\documentclass{desyproc}

\newcommand{\av}[1]{\left\langle #1 \right\rangle}

\newcommand{\gev}{\mathrm{GeV}}
\newcommand{\tev}{\mathrm{TeV}}

\newcommand{\mum}{\mathrm{\mu m}}

\newcommand{\pt}{p_{\rm t}}
\renewcommand{\d}{{\rm d}}

\newcommand{\Jpsi} {\mbox{J\kern-0.05em /\kern-0.05em$\psi$}\xspace}

\begin{document}

\title{ALICE first physics results}


%

%

%
\author{{\slshape Andrea Dainese}  for the ALICE Collaboration\\[1ex]
INFN - Sezione di Padova, via Marzolo 8, 35131 Padova,  Italy}

\contribID{xy}  
\confID{1964}
\desyproc{DESY-PROC-2010-01}
\acronym{PLHC2010}
\doi            

\maketitle

\begin{abstract}
ALICE is the dedicated heavy-ion experiment at the Large Hadron Collider.
The experiment has also a broad program of QCD measurements in proton--proton (pp) 
collisions, which have two-fold interest: the study of particle production at 
the highest energy frontier, and the definition of 
references for the corresponding measurements in the upcoming Pb--Pb run.
We present the first results on the pseudorapidity and
transverse-momentum dependence of charged 
particle production in pp collisions at LHC energies, on the $\rm \bar p/p$ ratio and on the 
Bose--Einstein particle correlations. As an outlook, we report on the status of the ongoing analyses for strangeness and heavy-flavour production measurements. 
\end{abstract}

\section{Introduction}
\label{sec:intro}

The ALICE experiment~\cite{aliceJINST,jurgen} will study nucleus--nucleus and 
proton--proton collisions at the Large Hadron Collider, with the
main goal of investigating the properties of the high-density 
state of QCD matter that is expected to be formed
in Pb--Pb collisions~\cite{PPRv1,PPRv2}.
The detector was designed in order to provide tracking and particle identification, for all particle species, over a large range of momenta
(from tens of MeV/$c$ to over 100~GeV/$c$), low material budget and excellent vertexing capabilities. These features have been tailored to reach a
detailed characterization of the state of matter produced in Pb--Pb collisions, with particular attention to global event
properties and hard probes. However, they also provide unique 
capabilities for carrying out a program of QCD measurements in pp collisions.

This report is organized as follows. In section~\ref{sec:ALICE}, the ALICE experimental setup is 
described, with emphasis on the detectors that were used for the results presented here,
along with the  data collection and event classification. The two most fundamental measurements
that characterize inclusive particle production are reported in sections~\ref{sec:mult} and \ref{sec:pt}: the charged particle multiplicity density and multiplicity distribution at $\sqrt{s}=0.9$, 
2.36 and 7~TeV~\cite{mult09,mult09236,mult7,peter}, and the charged particle transverse momentum ($\pt$) distribution and 
mean transverse momentum as a function of event multiplicity at 0.9~TeV~\cite{pt09,peter}.
In section~\ref{sec:pbarp} the measurement of the midrapidity 
antiproton over proton ratio,
which allows to address the mechanisms that transfer the baryon number from beam to central 
rapidity, at 0.9 and 7~TeV is described~\cite{pbarp097,michal}. In section~\ref{sec:femto} the measurement of the 
Bose-Einstein two-pion correlation, that allows to characterize the spatial extension 
of the particle emitting source, is described~\cite{femto09,dariusz}. Finally, in section~\ref{sec:strangecharm},
an outlook is given on the ongoing analyses on strangeness~\cite{antonin} 
and heavy-flavour production.

\section{ALICE detector, data collection and event classes}
\label{sec:ALICE}

The ALICE apparatus is described in~\cite{aliceJINST}. It consists of two main parts: 
a central detector,
placed inside a solenoidal magnet providing a field of up to 0.5~T, where charged and 
neutral particles
are reconstructed and identified in the pseudorapidity range $|\eta|<0.9$,
and a forward muon spectrometer covering the range $-4<\eta<-2.5$. The apparatus is completed
by a set of smaller detectors in the forward areas, for triggering, charged particle and photon 
counting, and event classification.

The main results presented in this report (sections~\ref{sec:mult}--\ref{sec:femto}) were obtained using the following ALICE detectors:
the VZERO scintillators, the Inner Tracking System (ITS), the Time Projection Chamber (TPC).

The two forward scintillator hodoscopes (VZERO) are segmented into 32 scintillator counters each, arranged in four rings around the beam pipe. They  cover the pseudorapidity ranges $2.8 < \eta < 5.1$ and $-3.7 < \eta < -1.7$, respectively. 

The ITS~\cite{marcello} is composed of high resolution silicon tracking detectors, arranged in six cylindrical layers at radial distances to the beam line from 3.9 to 43 cm. Three different technologies are employed:
Silicon Pixel Detectors (SPD) for the two innermost layers, Silicon Drift Detector (SDD) 
for the two intermediate layers, and Silicon Strip Detector (SSD) for the two outermost layers. 
The design spatial resolutions of the ITS sub-detectors ($\sigma_{r\phi}\times\sigma_z$) are: $12\times 100~\mum^2$ for SPD, $35\times 25~\mum^2$ for SDD, and $20\times 830~\mum^2$ for SSD. The SPD and SSD detectors were aligned using survey measurements, cosmic muon data~\cite{ITSalignCosm} and collision data to an estimated accuracy of $10~\mum$ for the SPD and $15~\mum$ for the SSD~\cite{marcello}. No alignment corrections are applied to the positions of the SDD modules, for which calibration and alignment are in progress. The estimated misalignment of the SDD modules is about $100~\mum$.

The TPC~\cite{tpcpaper,magnus} is a large cylindrical drift detector
with cathode pad readout multi-wire proportional chambers at the two edges.
 The active volume is $85 < r < 247$~cm and $-250 < z < 250$~cm in the radial and longitudinal directions respectively. 
At the present level of calibration, the transverse momentum resolution achieved in the TPC is given by $(\sigma_{\pt})/\pt)^2 =(0.01)^2+(0.007\,\pt)^2$, with $\pt$ in GeV/$c$. The transverse momentum resolution for $\pt	> 1~\gev/c$ is measured in cosmic muon events by comparing the muon momenta reconstructed in the upper and lower halves of the TPC~\cite{tpcpaper}. 
For $\pt	< 1~\gev/c$, 
the Monte Carlo estimate	of $\sigma(\pt)/\pt\simeq	1\%$	 was	cross-checked	using the measured $\rm K^0_S$ invariant mass distribution. The $\d E/\d x$ resolution is estimated to be about
$5\%$ for full-length tracks~\cite{tpcpaper}.

All data presented in this report were collected with a magnetic field of 0.5~T. The analyses
with pp collisions at $\sqrt{s}=0.9$ and $2.36~\tev$ are based on data collected in 
November and December 2009, while the analyses at $\sqrt{s}=7~\tev$ are based on data
collected in April and May 2010.
The data at 0.9 TeV and 7 TeV were collected with a trigger requiring a hit in the SPD or in either of the VZERO counters; i.e. essentially at least one charged particle anywhere in the 8 units of pseudorapidity. At 2.36 TeV, the VZERO detector was turned off; the trigger required at least one hit in the SPD ($|\eta| < 2$). The events were selected in coincidence with signals from two beam pick-up counters (BPTX), one on each side of the interaction region, indicating the passage of proton bunches. Control triggers taken (with the exception of the 2.36 TeV data) for various combinations of beam and empty-beam buckets were used to measure beam-induced and accidental backgrounds. Most backgrounds were removed as described in~\cite{mult09236}. The remaining background in the sample is typically of the order of $10^{-4}$ to $10^{-5}$ and can be neglected.

The total inelastic pp cross section is commonly subdivided into contributions from diffractive and non-diffractive processes. At 0.9~TeV, we perform our analyses for two classes of events: inelastic (INEL) and non-single-diffractive (NSD) pp collisions. The INEL sample is selected 
using the minimum-bias trigger condition described above (signal in SPD or in either of the VZERO counters). For the NSD analyses, a subset of this sample is selected offline by requiring a coincidence between the two VZERO detectors. This condition suppresses a significant fraction of the single-diffractive (SD) events.
The fractions of the different process types contributing to the selected event samples are 
estimated by a Monte Carlo simulation. 
The process fractions of single-diffractive and double-diffractive
(DD) events in the event generators are scaled to match the cross section in $\rm p\bar p$ at $\sqrt{s}=0.9~\tev$ measured by the UA5 experiment~\cite{ua5}. The selection efficiency for INEL and NSD events is approximately 96\% and 93\%, respectively.
Since the 2.36~TeV data sample was triggered by at least one hit in the SPD, 
this selection was used for both INEL and NSD analyses.
At 7 TeV, there is no experimental information available about diffractive processes; 
therefore, we chose an event class requiring at least one charged particle in the pseudorapidity interval $|\eta| < 1$ ($\rm INEL>0$), minimizing the model dependence of the corrections. 
For the comparison of the multiplicity measurements at all LHC energies, 
we analyzed the data at 0.9 TeV and 2.36 TeV also in this event class.

\section{Results on charged particle multiplicity at $\sqrt{s}= 0.9$, $2.36$, and $7~\tev$}
\label{sec:mult}

ALICE has measured the charged particle multiplicity density $\d N_{\rm ch}/\d\eta$
and the multiplicity distribution $\d N_{\rm events}/\d N_{\rm ch}$
at $\sqrt{s}= 0.9$, $2.36$, and $7~\tev$  in $|\eta|<1.3$ ($1.0$ at $7~\tev$)~\cite{mult09,mult09236,mult7}.
The analysis is based on using hits in the two SPD layers to form short track segments, called tracklets. A tracklet is defined by a hit combination, one hit in the inner and one in the outer SPD layer, pointing to the reconstructed vertex. The tracklet algorithm is described in~\cite{mult09,mult09236}.
For this analysis, the position of the interaction vertex is reconstructed by correlating hits in the two silicon-pixel layers~\cite{davide}. The vertex resolution achieved depends on the particle multiplicity, and is typically $100$--$300~\mum$ in the longitudinal ($z$) and 
$200$--$500~\mum$ in the transverse direction.
Primary charged particles are defined as the particles produced in the collision, excluding the weak decays of strange hadrons. Their multiplicity 
is estimated by counting the number of SPD tracklets, corrected for:  geometrical acceptance, detector and reconstruction efficiencies;
contamination from weak-decay products of strange particles, gamma conversions, and secondary interactions;
undetected particles below the 50~MeV/$c$ transverse-momentum cut-off, imposed by absorption in the material;
combinatorial background in tracklet reconstruction.
Two different event generators, PYTHIA~\cite{pythia} (tune Perugia-0~\cite{perugia0}) and PHOJET~\cite{phojet}, were used to evaluate the corrections, using the PYTHIA results
as central value and the PHOJET results to define an asymmetric systematic error.
Other systematic uncertainties were estimated as detailed in~\cite{mult09236}. 
The main error sources that were considered are: detector material description, SPD residual misalignment, particle composition in the generators, fraction of particle below the 
low-momentum cut-off, relative fraction of 
non-diffractive, single-diffractive and double-diffractive events. 

\begin{table}[!ht]
\centering
\begin{tabular}{|cccccc|}
  \hline
  Energy & ALICE  & \multicolumn{3}{c}{PYTHIA~\cite{pythia}} & PHOJET~\cite{phojet} \\
  \cline{3-5}
    (TeV)&        &  (109)~\cite{D6Ttune} &   (306)~\cite{CSCtune}   &   (320)~\cite{perugia0} &  \\
\hline
   & \multicolumn{5}{c|}{Charged-particle pseudorapidity density}\\
  \hline
  0.9    &   $3.81 \pm 0.01 ^{+0.07}_{-0.07}$ & 3.05 & 3.92 & 3.18 & 3.73 \\
  2.36   &   $4.70 \pm 0.01 ^{+0.11}_{-0.08}$ & 3.58 & 4.61 & 3.72 & 4.31 \\
  7      &   $6.01 \pm 0.01 ^{+0.20}_{-0.12}$ & 4.37 & 5.78 & 4.55 & 4.98 \\
\hline
   & \multicolumn{5}{c|}{Relative increase (\%)}\\
  \hline
  0.9--2.36 & $23.3 \pm 0.4 ^{+1.1}_{-0.7}$ & 17.3 & 17.6 & 17.3 & 15.4 \\
  0.9--7    & $57.6 \pm 0.4 ^{+3.6}_{-1.8}$ & 43.0 & 47.6 & 43.3 & 33.4 \\
\hline
\end{tabular}
\caption{\label{tab:multip}
$\d N_{\rm ch}/\d\eta$ at central pseudorapidity ($|\eta|<1$), for inelastic collisions having at least one charged particle in the same region (INEL$>0$), at three centre-of-mass energies~\cite{mult7}. For ALICE, the first uncertainty is statistical and the second is systematic. The relative increases between the 0.9~TeV and 2.36~TeV data, and between the 0.9~TeV and 7~TeV data, are given in percentages. The experimental measurements are compared to the predictions from models. For PYTHIA the tune versions are given in parentheses. The correspondence is as follows: D6T tune (109), ATLAS-CSC tune (306), and Perugia-0  tune (320).}
\end{table}

\begin{figure}[!t]
\centering
\includegraphics[width=0.55\textwidth]{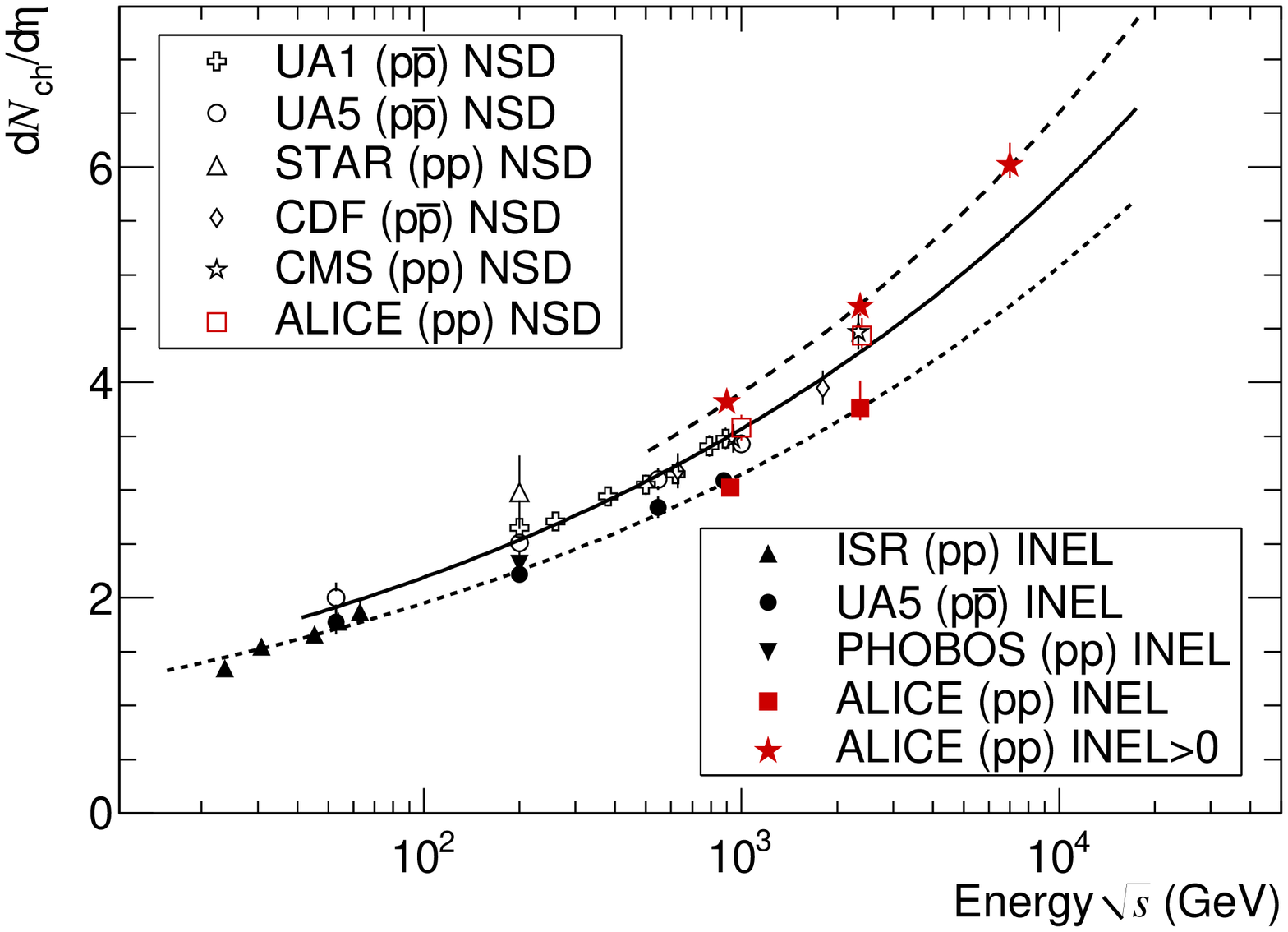}
\includegraphics[width=0.44\textwidth]{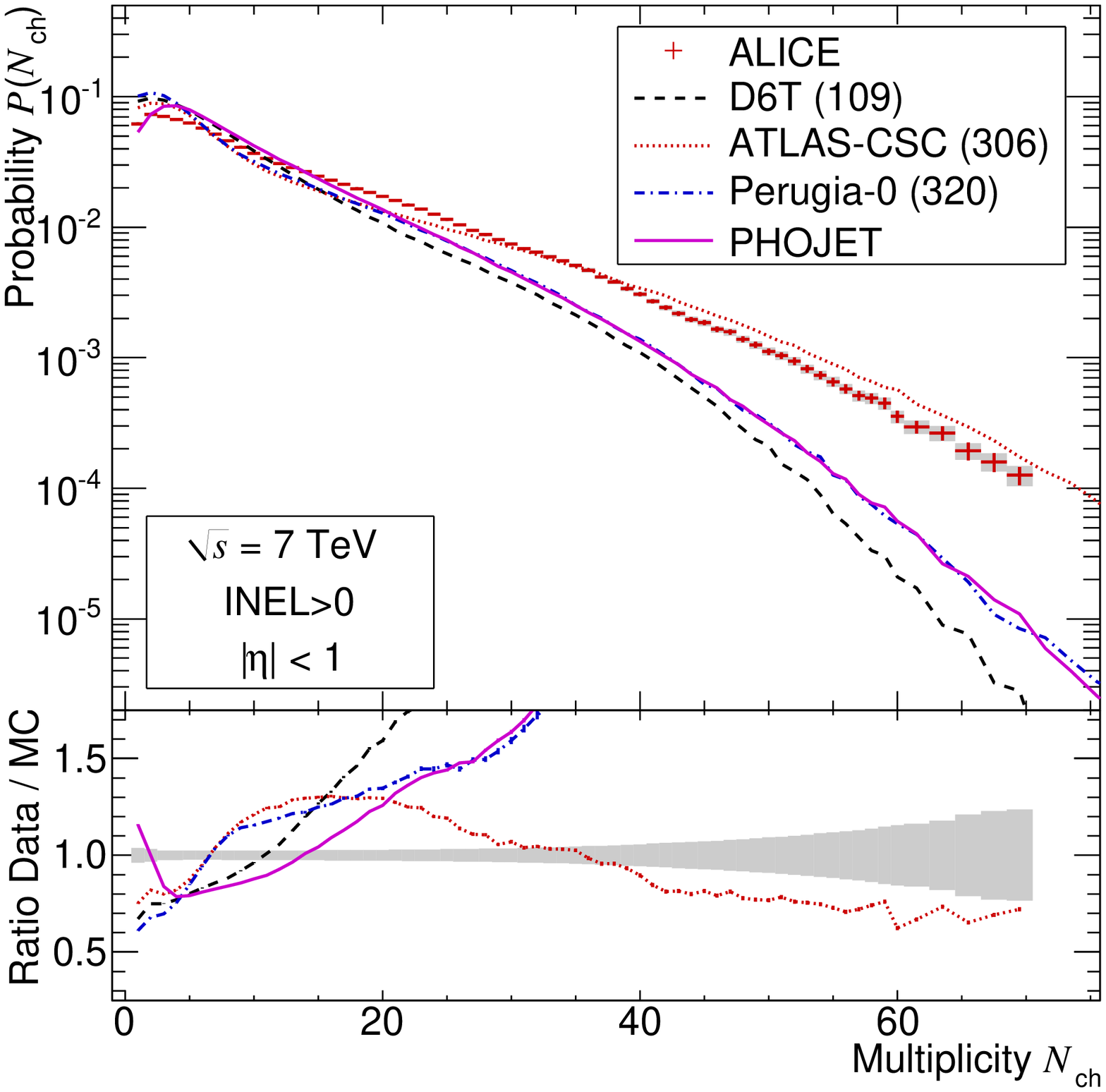}
\caption{Left: Charged-particle pseudorapidity density in the central pseudorapidity region $|\eta|<0.5$
 for inelastic and non-single-diffractive collisions, and in $|\eta|<1$ for inelastic collisions with at least one charged particle in that region (INEL$>$0), as a function of the centre-of-mass 
 energy~\cite{mult7}. The lines indicate the
fit using a power-law dependence on energy. 
Right: multiplicity distribution at 7~TeV in $|\eta|<1$ for the INEL$>$0 event class~\cite{mult7}. 
The error bars for data points represent statistical uncertainties, the shaded areas represent systematic uncertainties. The data are compared to models: PHOJET (solid line), PYTHIA tunes D6T (dashed line), ATLAS-CSC (dotted line) and Perugia-0 (dash-dotted line). In the lower part, the ratios between the measured values and model calculations are shown with the same convention. The shaded area represents the combined statistical and systematic uncertainties.}
\label{fig:mult}
\end{figure}

The pseudorapidity density of primary charged particles in the central pseudorapidity region 
$|\eta| < 1$ are presented in Table~\ref{tab:multip} and compared to models. The measured values are higher than those from the models considered, except for PYTHIA tune ATLAS-CSC~\cite{CSCtune} for the $0.9~\tev$ and $2.36~\tev$ data, and PHOJET for the 0.9~TeV data, which are consistent with the data. At 7~TeV, the data are significantly higher than the values from the models considered, with the exception of PYTHIA tune ATLAS-CSC, for which the data are only two standard deviations higher. We have also studied the relative increase of pseudorapidity densities of charged particles (Table~\ref{tab:multip}) between the measurement at 0.9~TeV and the 
measurements at 2.36~TeV and 7~TeV. We observe an increase
of $57.6\%\pm 0.4\%\,(stat.)^{+3.6}_{-1.8}\%\,(syst.)$ between the 0.9~TeV
and 7~TeV data, compared with an increase of 47.6\% obtained from the closest model, PYTHIA tune ATLAS-CSC. Therefore, the measured multiplicity density increases with increasing energy significantly faster than in any of the models considered.
In Fig.~\ref{fig:mult} (left), the centre-of-mass energy dependence of the pseudorapidity density of charged particles is shown for the $\rm INEL>0$, INEL and NSD classes.
 Note that $\rm INEL>0$ values are higher than inelastic and non-single-diffractive values, as expected, because events with no charged particles in $|\eta|< 1$ are removed. 
 The energy dependence is well described by a power-law with $\d N_{\rm ch}/\d\eta\propto \sqrt{s}^\alpha$ ($\alpha\simeq 0.2$) and extrapolates to the design LHC energy of $14~\tev$ with values that range from 5.7 for INEL to 7.4 for $\rm INEL>0$.

The multiplicity distributions $\d N_{\rm events}/\d N_{\rm ch}$ were measured at the three energies.
The raw measured distributions were corrected for efficiency, acceptance, and other detector effects, using a method based on unfolding with a detector response matrix from Monte Carlo simulations~\cite{mult09236}. The unfolding procedure applies $\chi^2$ minimization with regularization. 
The multiplicity distribution at 7~TeV is shown in Fig.~\ref{fig:mult} (right)
A comparison with models shows that only the PYTHIA tune ATLAS-CSC is close to the data at high multiplicities ($N_{\rm ch} > 25$). However, it does not reproduce the data in the intermediate multiplicity region ($8 < N_{\rm ch} < 25$). At low multiplicities, ($N_{\rm ch} < 5$), there is a large spread of values between different models: PHOJET is the lowest and PYTHIA tune Perugia-0 the highest. Similar comparisons for 0.9 and 2.36~TeV can be found in~\cite{mult09236}.

\section{Results on charged particle $\pt$ spectra at $\sqrt{s}= 0.9~\tev$}
\label{sec:pt}

Charged particle tracks are reconstructed using information from the TPC and ITS detector systems. Signals on adjacent pads in the TPC are connected to particle tracks by employing a Kalman filter algorithm. The TPC tracks are extrapolated to the ITS and matching hits in the ITS detector layers are assigned to the track.
The event vertex is reconstructed using the combined track information from TPC and ITS, and the measured average intersection profile as a constraint~\cite{davide}. 
The study of the transverse momentum spectrum of charged particles in pp at $\sqrt{s}=0.9~\tev$
is reported in~\cite{pt09}.
Tracks are selected in the pseudorapidity range $|\eta| < 0.8$. 
Additional quality requirements are applied to
ensure high tracking resolution and low secondary and
fake track contamination. A track is accepted if it has
at least 70 out of the maximum of 159 space points in
the TPC, and the $\chi^2$ per space point used for the momentum fit is less than 4. 
Additionally, at least two
hits in the ITS must be associated with the track, and
at least one has to be in either of the two innermost layers, i.e., in the SPD.  
Tracks with $\pt	< 0.15~\gev/c$ are excluded because their reconstruction efficiency drops below 50\%. Tracks are also
rejected as not associated to the primary vertex if their
distance of closest approach to the reconstructed event
vertex in the plane perpendicular to the beam axis, $d_0$,
satisfies $|d_0| > (350 + 420\,\pt^{-0.9})~\mum$, with $\pt$ in GeV/$c$. This cut corresponds to about seven standard deviations of the $\pt$-dependent transverse impact parameter 
resolution for primary tracks passing the above selection. 
The primary charged particle track reconstruction efficiency is about 75\% for $\pt> 0.6~\gev/c$. Below this $\pt$, the efficiency decreases and reaches 50\% at $0.15~\gev/c$. 
The contamination from secondary particles is 9\% at 0.15~GeV/$c$ and 
and drops below 3\% for above $1~\gev/c$~\cite{pt09}.
The reconstruction efficiency and contamination, evaluated with the PYTHIA event generator, are converted to $\pt$ dependent correction factors used to correct the raw $\pt$ spectrum.
For the normalization of the transverse momentum spectra to the number of events, multiplicity dependent correction factors are derived from the event selection and vertex reconstruction efficiencies for INEL and NSD events, evaluated with the PYTHIA event generator.

\begin{figure}[!t]
\centering
\includegraphics[width=0.4\textwidth]{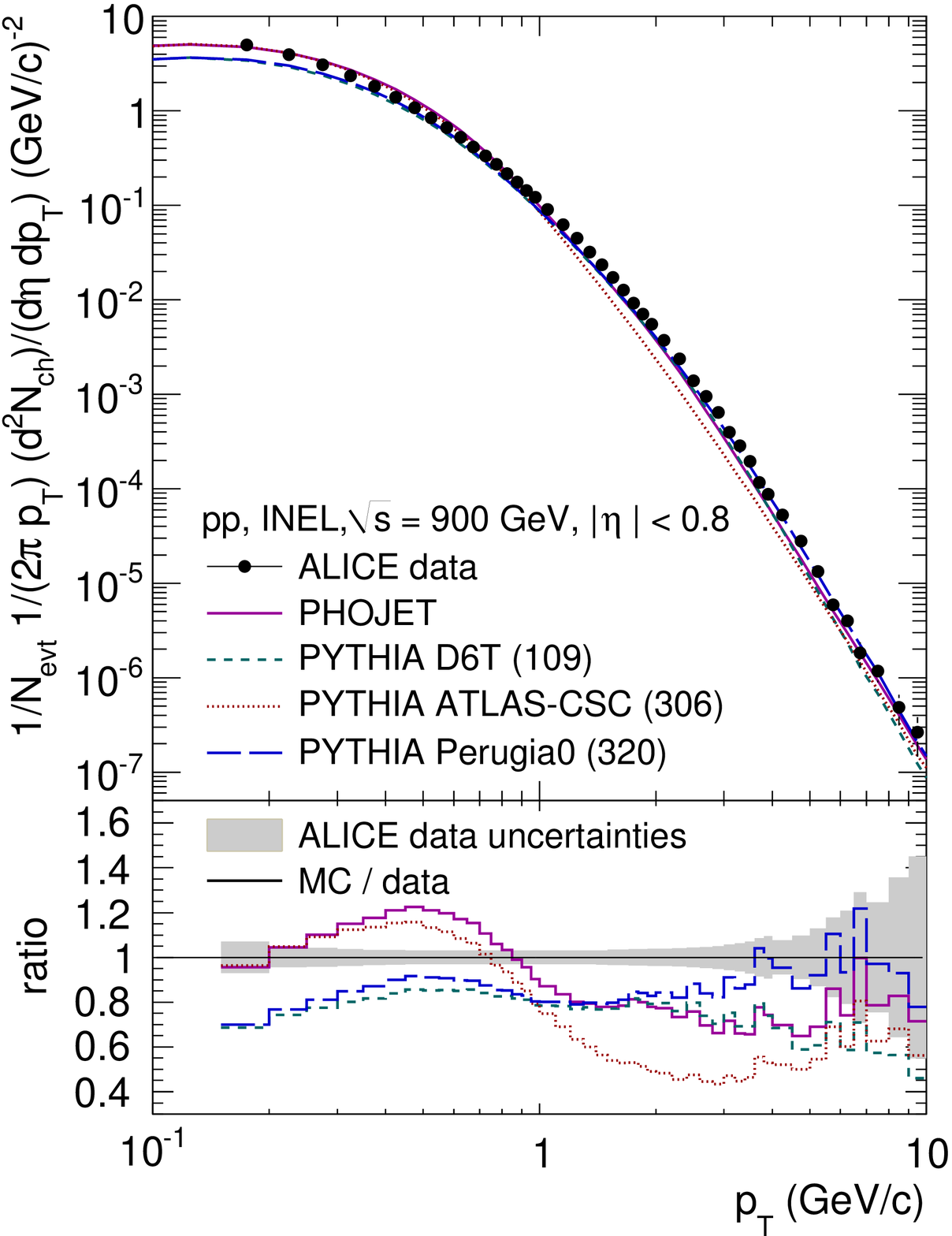}
\includegraphics[width=0.4\textwidth]{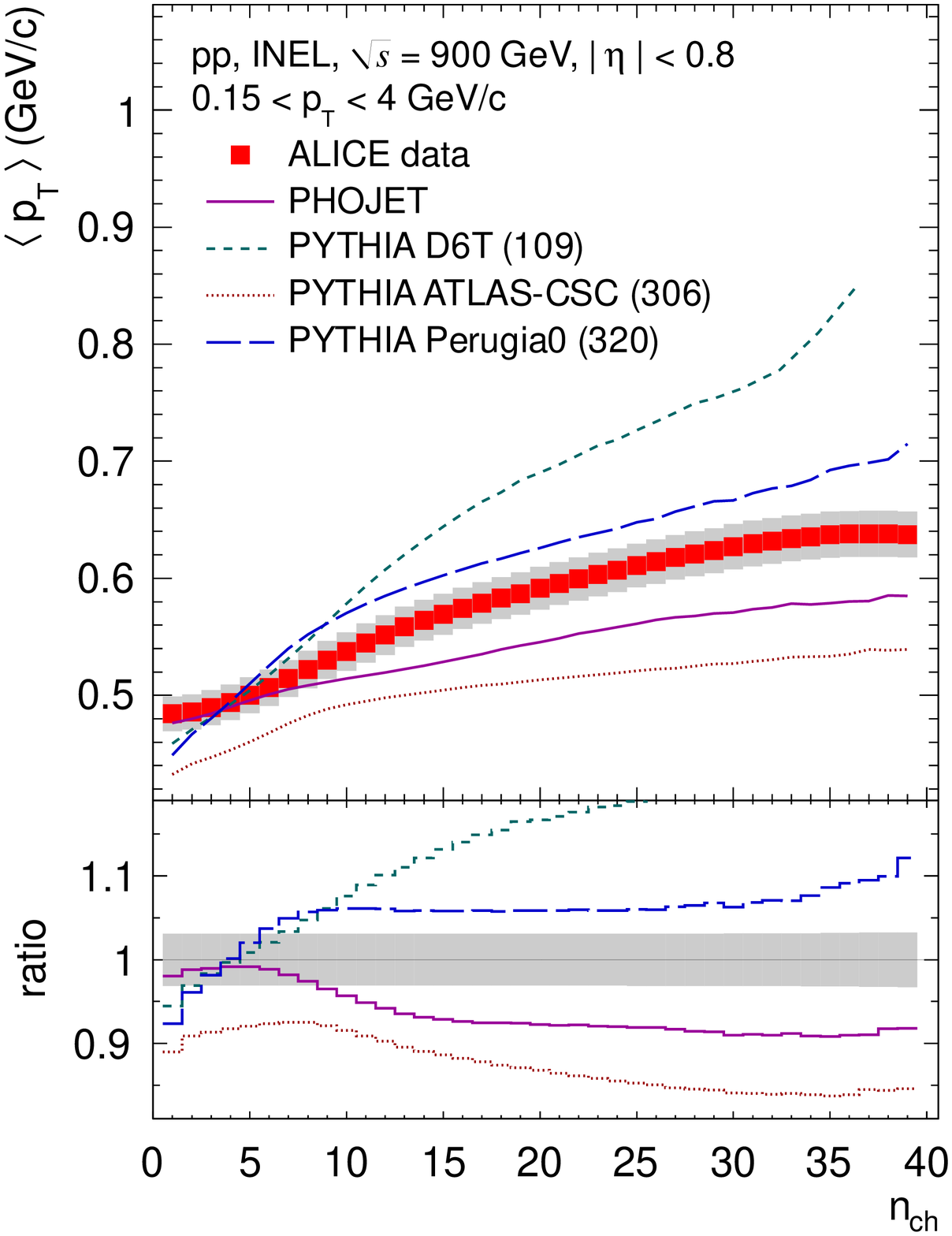}
\caption{Left: primary charged particle 
$\pt$-differential yield in INEL pp collisions at $\sqrt{s}=900$~GeV 
($|\eta|<0.8$), compared to results from PHOJET and PYTHIA tunes D6T~\cite{D6Ttune},
ATLAS-CSC~\cite{CSCtune} and Perugia-0~\cite{perugia0}. 
Right: The average transverse momentum of charged particles for $0.15<p_T<4$~GeV/$c$
as a function of $n_{\rm ch}$, in comparison to models. The error bars and the shaded area
indicate the statistical and systematic errors of the data, respectively.
In the lower panels, the ratio Monte Carlo over data is shown. 
The shaded areas 
indicate the statistical and systematic uncertainty of the data,
added in quadrature. Figures from~\cite{pt09}.}
\label{fig:pt}
\end{figure}

In Fig.~\ref{fig:pt} the results on $(1/2\pi\pt){\rm d}^2N_{\rm ch}/{\rm d}\eta{\rm d}\pt$ for INEL pp events at $\sqrt{s}=0.9~\tev$~\cite{pt09} are shown and 
compared to PHOJET and different tunes of PYTHIA, D6T (tune 109), Perugia-0 (tune 320)and ATLAS-CSC (tune 306). The best agreement is found with the Perugia-0 tune, which gives a fair description of the spectral shape, but is approximately 20\% below the data. 
The D6T tune is similar to Perugia-0 below $2~\gev/c$ but underestimates the data more significantly at high $\pt$. PHOJET and the PYTHIA ATLAS-CSC tune fail to reproduce 
the spectral shape of the data.
We note that PHOJET and ATLAS-CSC agree best with the charged particle multiplicity measurements at $\sqrt{s} = 0.9$ and 2.36, and 7 TeV, respectively (see Table~\ref{tab:multip}).

The average transverse momentum $\av{\pt}$  (in the range $0.15<\pt<4~\gev/c$) 
as a function of the acceptance
and efficiency corrected multiplicity ($n_{\rm ch}$) is shown in the right-hand panel of Fig.~\ref{fig:pt}
(see~\cite{pt09} for analysis details). A significant increase of $\av{\pt}$ with multiplicity is observed. 
Event generator curves are also shown and indicate that
Perugia-0 and PHOJET are the closest to the data, however, none of the models gives a good description of the entire measurements.

The analysis of the transverse momentum spectra of charged particles at 7~TeV, 
currently ongoing, will allow to extend the $\pt$ reach to about $50~\gev/c$.
Other $\pt$ spectra analyses in progress include the identified charged hadrons 
($\pi$, K, and p),
using the PID capabilities of the ITS, TPC and TOF detectors, and the neutral mesons ($\pi^0$
and $\eta$), using photon pairs reconstructed via $\gamma\to e^+e^-$ conversions in the material 
as well as via the two ALICE electromagnetic calorimeters, PHOS and EMCAL.

\section{Results on $\rm \bar p/p$ ratio at $\sqrt{s}= 0.9$ and $7~\tev$}
\label{sec:pbarp}

The $\rm \bar p/p$ ratio was measured in pp collisions at $\sqrt{s}= 0.9$ and $7~\tev$ in 
the ranges $|y|<0.5$ and $0.45<\pt<1~\gev/c$~\cite{pbarp097}. The physics motivation for this
measurement is the study of the baryon transport mechanism over large rapidity intervals 
in high-energy proton--proton collisions.
In inelastic non-diffractive proton-proton collisions at very high
energy, the conserved baryon number associated with the
beam particles is often called {\em baryon-number transport} and has been
debated theoretically for some time (see references in~\cite{pbarp097}).
This 
baryon-number transport is usually quantified in terms of the rapidity loss
$\Delta y = y_{\rm{beam}} - y_{\rm{baryon}}$, where $y_{\rm{beam}}$
($y_{\rm{baryon}}$) is the rapidity of the incoming beam (outgoing 
baryon).
The LHC opens the
possibility to investigate baryon transport over very large rapidity intervals ($\Delta y=6.9$ and 8.9 
at $\sqrt{s}=0.9$ and 7~TeV, respectively)
by measuring the antiproton-to-proton production ratio at midrapidity,
$R = N_{\overline{\rm p}}/N_{\rm p}$.
Most of the protons and antiprotons at midrapidity are created in baryon--antibaryon
pair production, implying equal yields. Any excess of protons over antiprotons
is therefore associated with the baryon-number transfer from the incoming
beam. Model predictions for the ratio 
$R$ at LHC energies range from unity, i.e. no baryon-number transfer to 
midrapidity, in models where the baryon-number transfer is suppressed 
exponentially with the rapidity interval $\Delta y$, down to about 0.9, in models where the baryon-number transfer does not depend on $\Delta y$.

For the analysis, the track selection described in Section~\ref{sec:pt} was used.
Protons were identified using their
$\d E/\d x$ signal in the TPC. In the restricted acceptance defined by 
$|y|<0.5$ and $0.45<p<1.05~\gev/c$, the residual contamination
from other hadrons and leptons is $<0.1\%$.
For the rejection 
of secondary protons from strange baryon decays, 
a $\pt$-dependent impact parameter cut was used, specifically optimized for protons, which 
are detected with poorer resolution than pions.  The residual secondary contamination 
was measured from the data, using the impact parameter distributions~\cite{pbarp097}.
Since the aim of the analysis is a sensitivity on $R$ of order $1\%$, special attention was
placed on the evaluation of the acceptance and efficiency corrections, and in particular 
on the corrections for proton and antiproton elastic and inelastic (absorption) in the
detector material. This was done comparing the cross sections for these processes
in different particle transport models and with existing data~\cite{pbarp097}.

\begin{figure}[!t]
\includegraphics[width=0.37\textwidth]{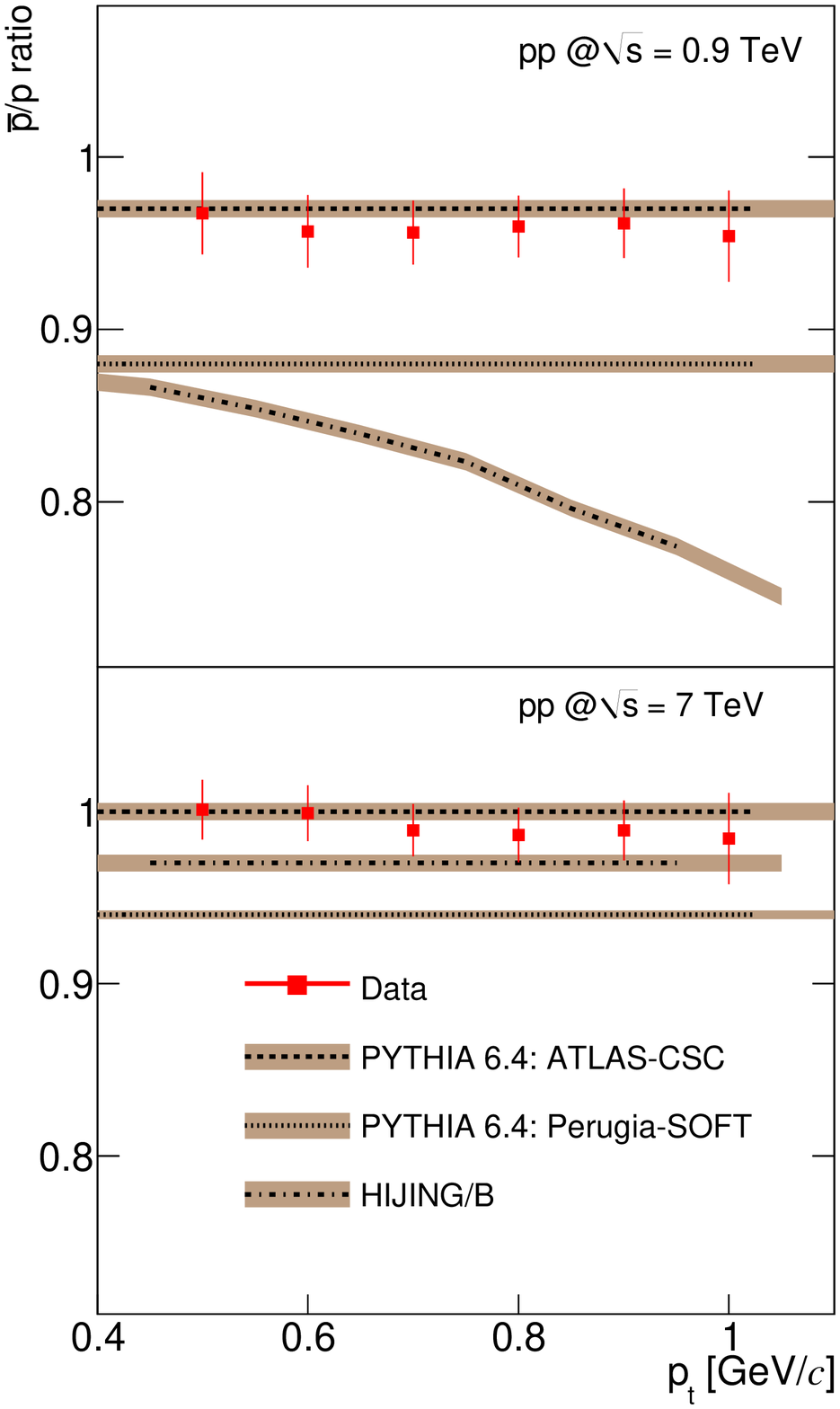}
\hfill
\includegraphics[width=0.53\textwidth]{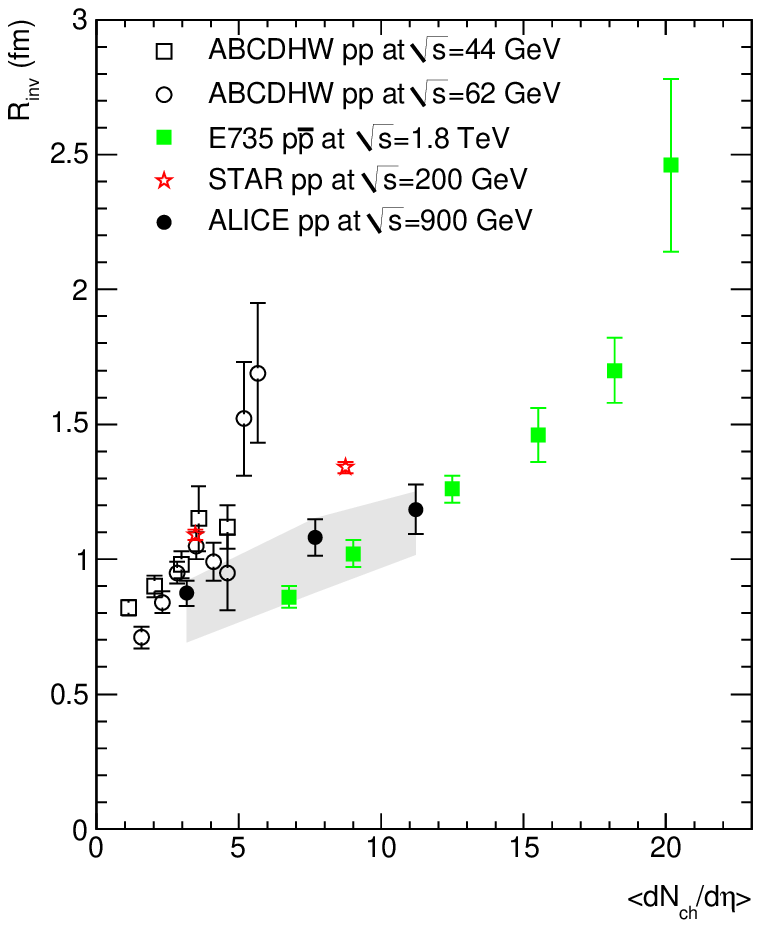}
\caption{Left: $\rm \bar p/p$ ratio as a function of $\pt$ in $|y| <0.5$ for pp 
at $\sqrt{s} = 0.9$~TeV (top) and
7~TeV (bottom)~\cite{pbarp097}. Only statistical errors are shown for the data;
the width of the Monte Carlo bands indicates the statistical uncertainty of 
the simulation results.
Right: Bose-Einstein correlation Gaussian radius, as a function of the charged-particle multiplicity at midrapidity (full dots), in pp at $\sqrt{s} = 0.9$~TeV~\cite{femto09}. 
The shaded band represents the systematic errors. 
For comparison, the data taken at the ISR, RHIC, and Tevatron, are shown (see references 
in~\cite{femto09}).}
\label{fig:pbarpandfemto}
\end{figure}

The final $\rm \bar p/p$ ratio $R$ integrated within our rapidity and $\pt$ acceptance rises from
$R = 0.957\pm 0.006(stat.)\pm 0.014(syst.)$ at $\sqrt{s}=0.9$~TeV to $R= 0.991\pm0.005(stat.) \pm 0.014(syst.)$ at $\sqrt{s} = 7$~TeV~\cite{pbarp097}. The difference in the ratio, $0.034\pm
0.008(stat.)$ is significant because the systematic errors at both energies are fully correlated.
Within statistical errors, the measured ratio $R$ shows no dependence on transverse momentum (see left panel of Fig.~\ref{fig:pbarpandfemto}) or rapidity (data not shown).
Our measurement is compatible with $R=1$ at the highest LHC energy, thus excluding mechanisms that do not suppress the baryon-number transport over large $\Delta y$.
Indeed, as seen in Fig.~\ref{fig:pbarpandfemto} (left), the models that implement these
mechanisms, PYTHIA with Perugia-SOFT tune and HIJING/B, underpredict our result.

\section{Results on Bose-Einstein correlations at $\sqrt{s}= 0.9~\tev$}
\label{sec:femto}

Bose-Einstein enhancement of identical-pion pairs at low relative momentum allow 
to assess the spatial scale of the emitting source in $e^+e^-$, hadron--hadron, lepton--hadron, 
and heavy-ion collisions. Especially in the latter case, this technique, known as 
Hanbury Brown-Twiss (HBT) interferometry and being a special case of femtoscopy, has been developed into a precision tool to probe the dynamically-generated geometry of the emitting system. See~\cite{femto09} for more details and references. 
A systematic program of femtoscopic measurements in pp and heavy-ion collisions at the LHC 
will allow to investigate the nature, the similarities, and the differences of their dynamics. 
This program was started by measuring the two-pion correlations in pp collisions at 
$\sqrt{s}=0.9~\tev$~\cite{femto09}. Pions tracks are reconstructed in the TPC and ITS 
(similar selection cuts as for the $\pt$ spectrum analysis) and identified using the TPC
$\d E/\d x$. 
The analysis of the correlation function (details in~\cite{femto09}) shows an increase of the extracted radius of the correlation volume with increasing event multiplicity, in line with other measurements done in particle and nuclear collisions, see Fig.~\ref{fig:pbarpandfemto} (right). Conversely, the strong decrease of the radius with increasing pair transverse momentum, as observed at RHIC and at Tevatron, is not manifest in our data (not shown here, see~\cite{femto09}).

\section{Prospects for strangeness and charm production measurements at $\sqrt{s}= 7~\tev$}
\label{sec:strangecharm}

\begin{figure}[!b]
\includegraphics[width=0.49\textwidth]{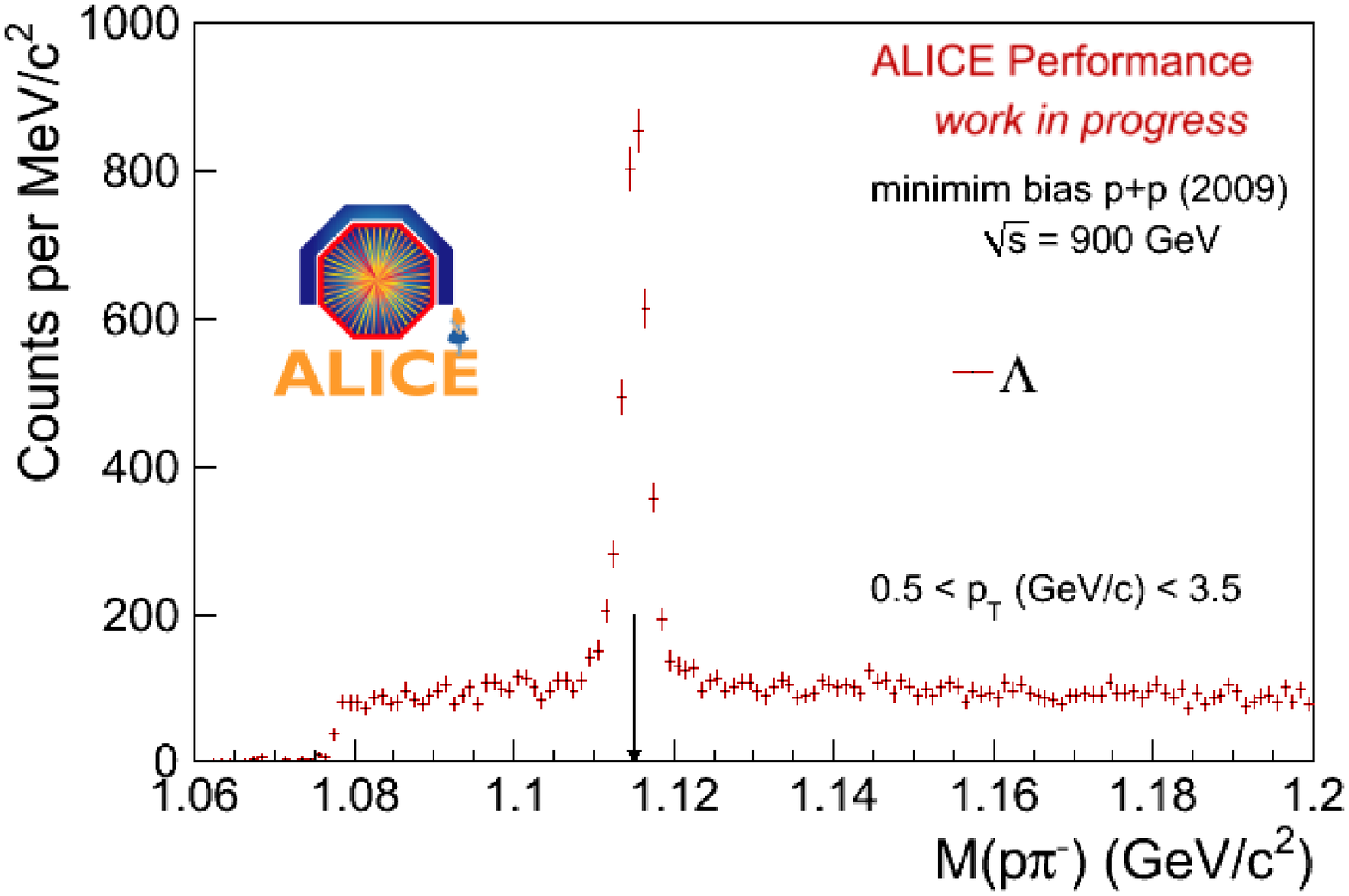}
\includegraphics[width=0.49\textwidth]{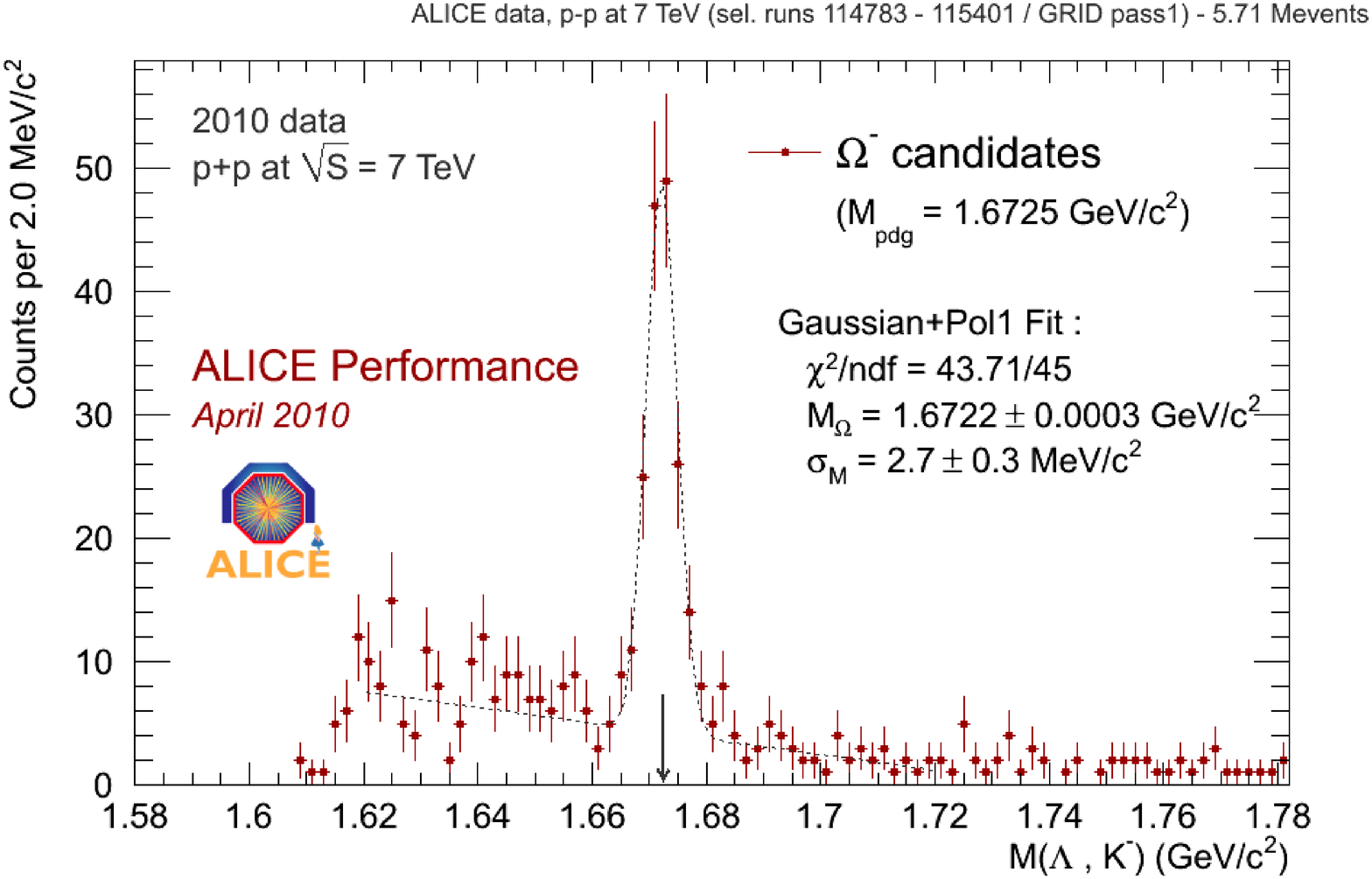}
\includegraphics[width=0.42\textwidth]{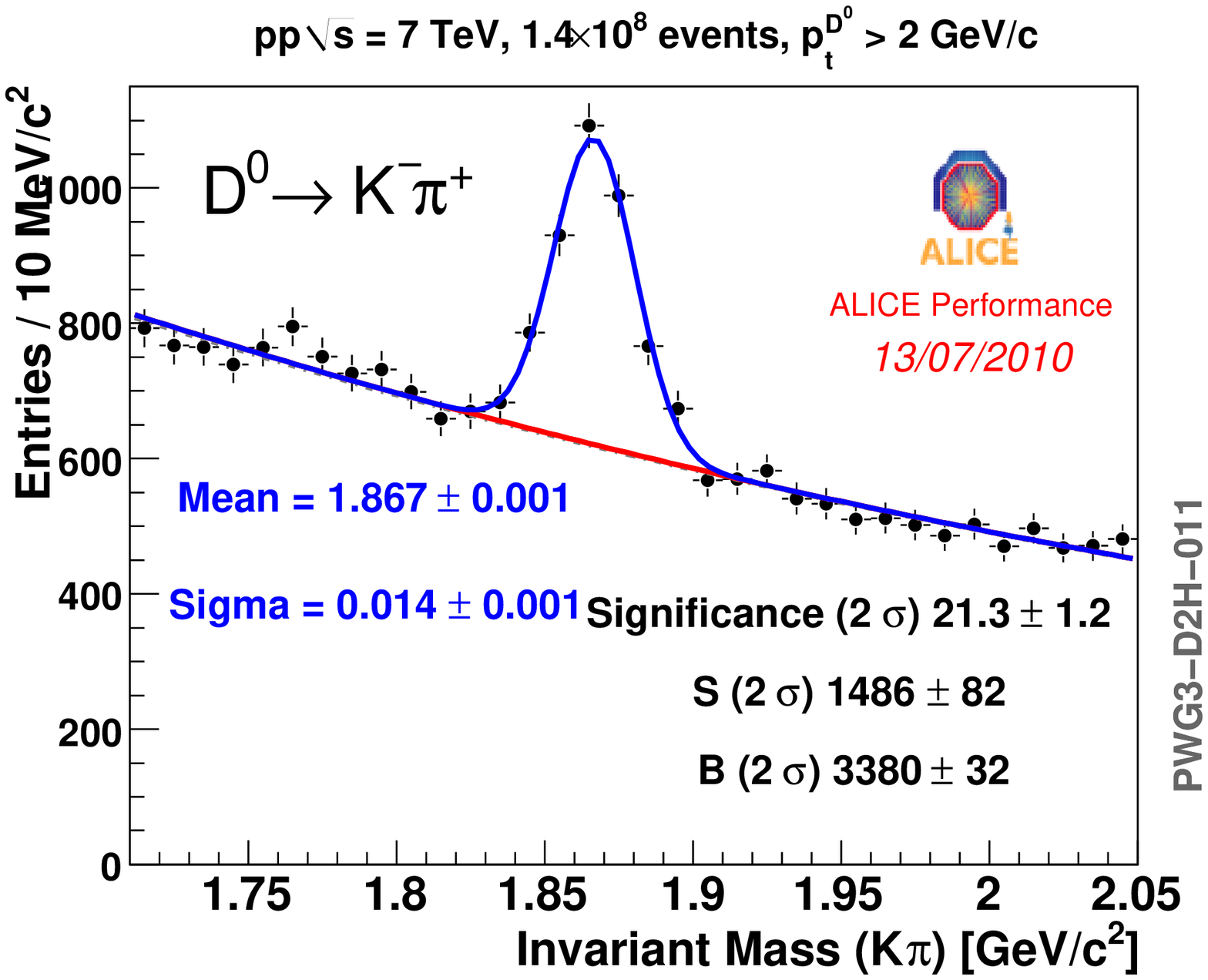}
\hfill
\includegraphics[width=0.49\textwidth]{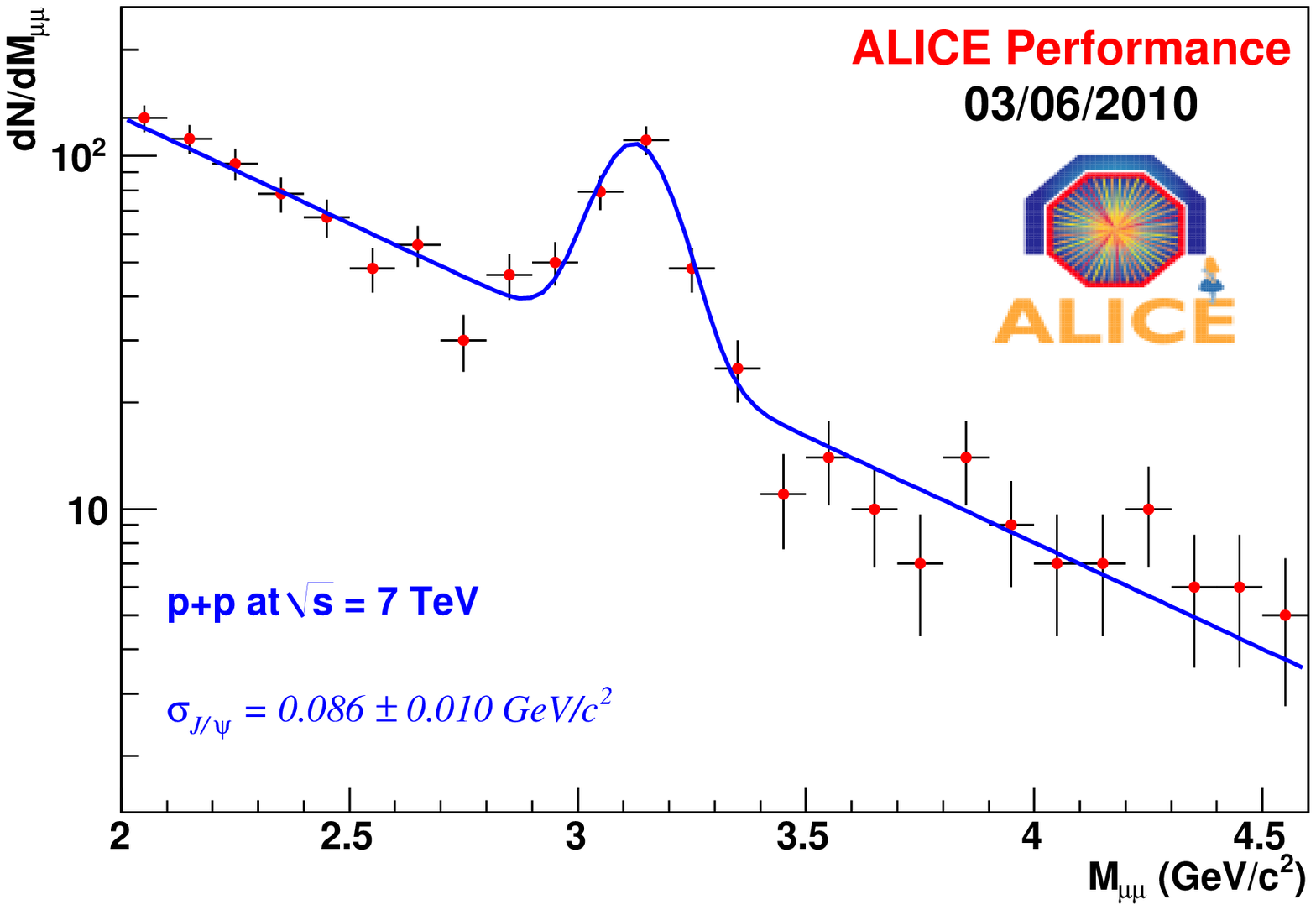}
\caption{Example signals (invariant mass distributions) for 
$\rm \Lambda\to p\pi^-$ at 0.9~TeV (top-left), $\rm \Omega^-\to \Lambda K^-$ (top-right),
$\rm D^0\to K^-\pi^+$ (bottom-left) and $\rm J/\psi\to\mu^+\mu^-$ (bottom-right) at 7~TeV.}
\label{fig:masses}
\end{figure} 

Several measurements of strange and heavy-flavour particle production are being prepared,
using pp collision data at 0.9 and 7~TeV.

In particular, the following strange mesons and baryons are reconstructed topologically 
in ALICE: $\rm K^0_S$,
$\rm K^{*0}$, $\phi$, $\Lambda$, $\Xi^-$, $\Omega^-$, $\Sigma^{*-}$. As examples,
in the upper panels of Fig.~\ref{fig:masses}, we show the signals for 
$\Lambda$ at 0.9~TeV and $\Omega^-$ at 7~TeV.
 
Charm and beauty production measurements are in preparation using: 
at central rapidity, hadronic decays of D mesons ($\rm D^0\to K^-\pi^+$, 
$\rm D^0\to K^-\pi^+\pi^-\pi^+$, $\rm D^{*+}\to D^0\pi^+$, $\rm D^+\to K^-\pi^+\pi^+$, $\rm D_s^+\to K^-K^+\pi^+$) and single electrons from D and B semi-electronic decays, identified in the TPC, TOF, Transition Radiation Detector (TRD) and EMCAL; at forward rapidity, 
single muons and di-muons from $\rm c\bar c$ and $\rm b\bar b$.
Quarkonia will be reconstructed at central rapidity using di-electrons and at
forward rapidity using di-muons. $\rm J/\psi$ signals are already well visible in the di-electron
($|y|<1$) and di-muon ($-4<y<-2.5$) invariant mass distributions for pp collisions at $\sqrt{s}=7~\tev$.
In the lower panels of Fig.~\ref{fig:masses}, we show example signals for 
$\rm D^0\to K^-\pi^+$ and $\rm J/\psi\to\mu^+\mu^-$ at 7~TeV.

 \section{Summary}
 \label{sec:summary}

We have presented the first ALICE physics results for pp collisions at LHC:
\begin{itemize}
\item particle multiplicity at LHC increases with $\sqrt{s}$ energy significantly faster
than predicted by all models;
\item the mean transverse momentum evolution with event multiplicity at 0.9~TeV
is not reproduced by any of the commonly used event generator tunes;
\item the net baryon number at midrapity goes to unity at 7~TeV, implying that 
baryon number transfer over large rapidity intervals is suppressed;
\item the Bose-Einstein femtoscopic measurement show that the size of the 
correlation volume for particle production increases with event multiplicity.
\end{itemize}
Many other analyses are ongoing, as we have shown with some examples on 
strangeness and charm production.
ALICE has just started to deliver physics results and looks forward to the 
imminent LHC heavy-ion run~\cite{tori}.
 

\begin{footnotesize}

\end{footnotesize}



\begin{thebibliography}{99}

\bibitem{aliceJINST} 
 K.~Aamodt {\it et al.} [ALICE coll.], J. Instrum \textbf{3} (2008) S08002.
 
 \bibitem{jurgen} J. Schukraft [ALICE coll.], {\it these proceedings}.
 
 \bibitem{PPRv1}
  B.~Alessandro {\it et al.}  [ALICE coll.],
  J.\ Phys.\ G {\bf 32} (2006) 1295.
 
 \bibitem{PPRv2}
  F.~Carminati {\it et al.}  [ALICE coll.],
  J.\ Phys.\ G {\bf 30} (2004) 1517.

\bibitem{mult09} K.~Aamodt {\it et al.} [ALICE coll.], Eur. Phys. J. C \textbf{65} (2010) 111.   

 \bibitem{mult09236}
 K.~Aamodt {\it et al.} [ALICE coll.], Eur. Phys. J. C \textbf{68} (2010) 89.


\bibitem{mult7} K. Aamodt {\it et al.} [ALICE coll.], arXiv:1004.3514v1 (2010).

 \bibitem{peter} P. Hristov [ALICE coll.], {\it these proceedings}.

\bibitem{pt09}
  K.~Aamodt {\it et al.}  [ALICE coll.],
  arXiv:1007.0719v1 (2010).

\bibitem{pbarp097}
  K.~Aamodt {\it et al.}  [ALICE coll.],
  arXiv:1006.5432v1 (2010).

\bibitem{michal} M. Broz [ALICE coll.], {\it these proceedings}.

\bibitem{femto09}
  K.~Aamodt {\it et al.}  [ALICE coll.],
  arXiv:1007.0516v1 (2010).

 \bibitem{dariusz} D. Miskowiec [ALICE coll.], {\it these proceedings}.

 \bibitem{antonin} A. Maire [ALICE coll.], {\it these proceedings}.
 
\bibitem{ITSalignCosm} 
 K.~Aamodt {\it et al.} [ALICE coll.], J. Instrum \textbf{5} (2010) P03003.

\bibitem{marcello} M. Lunardon, {\it these proceedings}.
 
 \bibitem{tpcpaper}
  J. Alme {\it et al.}, arXiv:1001.1950 (2010).

\bibitem{magnus} M. Mager, {\it these proceedings}.
  
\bibitem{ua5} 
   G.J.~Alner {\it et al.} [UA5 coll.], Z.~Phys.~C \textbf{33} (1986) 1.
   
\bibitem{davide} D. Caffarri [ALICE coll.], {\it these proceedings}.

\bibitem{pythia}
  T. Sj\"{o}strand, S.~Mrenna, P.~Skands, J. High Energy Phys. \textbf{2006} (2006) 05 026.

\bibitem{perugia0} P.Z.~Skands, arXiv:0905.3418 (2009).

\bibitem{phojet}
R.~Engel, J.~Ranft, S.~Roesler, Phys. Rev. D \textbf{52} (1995) 1459.

\bibitem{D6Ttune} M.G.~Albrow {\it et al.}, arXiv:hep-ph/0610012 (2006).

\bibitem{CSCtune} A.~Moraes, ATLAS Note ATL-COM-PHYS-2009-119 (2009).

\bibitem{tori} H. Torii, {\it these proceedings}.

\end{thebibliography}
\end{document}